\documentclass[11pt]{article}

\usepackage[utf8]{inputenc}
\usepackage[T1]{fontenc}
\usepackage[english]{babel}
\usepackage{lmodern}
\usepackage{microtype}
\usepackage{graphicx}
\usepackage{amsmath}
\usepackage[hidelinks]{hyperref}
\usepackage{geometry}
\usepackage{indentfirst}
\title{\textbf{Stop Hand-Holding Your Coding Agent:\\ Engineering the Loops that Replace Step-by-Step Prompting}}
\author{Sandeco Macedo\\ Instituto Federal de Goi\'as (IFG), Brazil\\ \texttt{sanderson.macedo@ifg.edu.br}\\ ORCID: \href{https://orcid.org/0000-0002-5255-596X}{0000-0002-5255-596X}}
\date{}

\begin{document}
\maketitle

\begin{abstract}
\noindent In mid-2026 a slogan reorganized how practitioners talk about coding agents: stop prompting your agent, start designing the loop that prompts it. We take this claim seriously and give it a careful treatment. We call the object of the new practice the \emph{loop specification}: a bounded, reusable artifact, made of a trigger, a goal, a verification step, a stopping rule and a memory, that a human hands to an agent harness (such as Claude Code or Codex) so the agent pursues a goal on its own, in place of step-by-step prompting. We distinguish this external loop specification from two things it is often confused with: an ordinary programming loop, and the internal perceive-act-observe cycle that the harness already provides as plumbing. We position loop engineering as a new layer in the progression from prompt to context to harness to loop, and we argue, against the stronger headlines, that it does not retire prompt engineering; loop and prompt are distinct tools with distinct uses. We offer four contributions: a definition and scope for the discipline; an anatomy and taxonomy of loop specifications organized around trigger, goal type, a five-level verification ladder, architecture, and named terminal states; a descriptive analysis of the Loop Library, a public corpus of fifty real loops that we code by hand; and a set of design principles and anti-patterns grounded in the scientific literature on self-correction, reward hacking and model-as-judge fragility. The corpus shows that practice has matured most where the discipline says it matters: seventy percent of loops verify in the autonomous zone of the ladder and seventy-four percent name their terminal states, while automated triggering and durable memory remain comparatively underdeveloped. We close with the limits the practice must respect, including the verification burden, comprehension debt and the risk of cognitive surrender.

\end{abstract}

\section{Introduction}
In the second week of June 2026, a single idea reorganized the discourse around coding agents: stop prompting the agent at every step and instead design the loop that prompts it. Practitioners building these tools stated the shift bluntly, from ``I don't prompt Claude anymore. I have loops running that prompt Claude'' to the sharper imperative, circulated to several million viewers in a day, that one should no longer be prompting coding agents but designing the loops that prompt them.\footnote{Quoted in The New Stack, ``Loop Engineering,'' \url{https://thenewstack.io/loop-engineering/}, and in A. Osmani, ``Loop Engineering,'' \url{https://addyosmani.com/blog/loop-engineering/}.} Around the same time, a public catalogue of reusable loops, the Loop Library, turned the slogan into concrete artifacts.\footnote{Loop Library (Forward Future), \url{https://signals.forwardfuture.ai/loop-library/}.}

The claim is striking and the practice is real, but the concept arrived through threads, talks and blog posts rather than through any reviewable account. That gap motivates this paper. We ask what exactly is being built, how it relates to the loops that already exist inside an agent, what a corpus of real loops actually looks like, and which of the practitioner claims survive contact with the scientific literature.

The first task is to fix the object of study, because the word ``loop'' carries at least three meanings. An ordinary programming loop is plain control flow. The internal cycle of an agent, the model running tools in a \texttt{while} over a stop condition, is the perceive-act-observe machinery that frameworks from ReAct \cite{yao2023react} onward formalized; it is part of the harness and it exists whether or not anyone designs it. Neither of these is what the new practice means. The \emph{loop specification} is a third thing: an external, bounded, reusable artifact, a trigger plus a goal plus a verification step plus a stopping rule plus a memory, that a human designs and hands to a harness such as Claude Code or Codex so the agent finds the work, does it, checks its own result, and knows when to stop or call for help. The harness supplies the engine; loop engineering writes the pilot.

\begin{figure}[h!]
\centering
\includegraphics[width=0.92\textwidth]{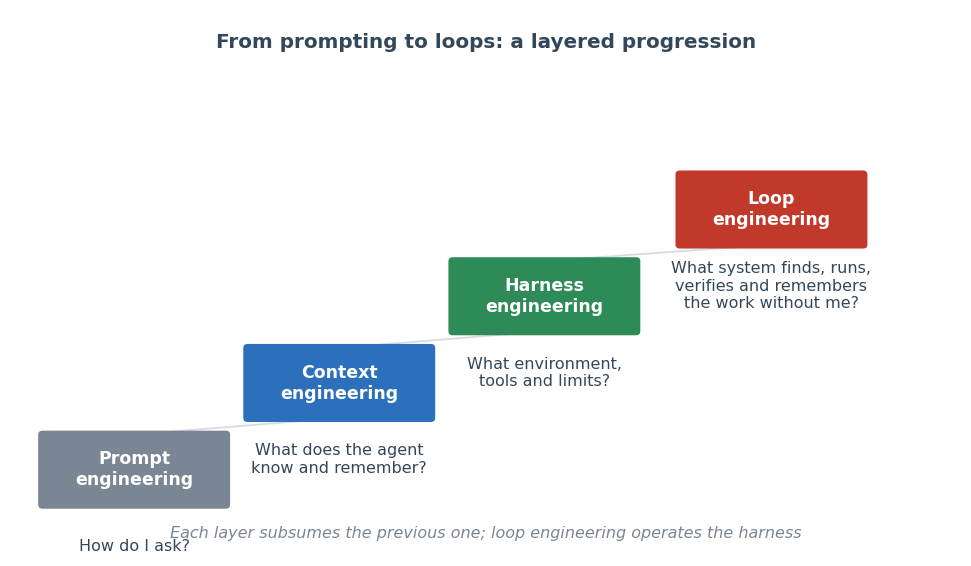}
\caption{Loop engineering as a new layer in the progression from prompt to context to harness to loop. Each layer subsumes the previous one; the loop specification operates the harness.}
\label{fig:progression}
\end{figure}

We situate the practice as a new layer in a progression (Figure~\ref{fig:progression}).\footnote{The four-layer framing is drawn from practitioner accounts of the discipline; see tosea.ai, ``What Is Loop Engineering?,'' \url{https://tosea.ai/blog/loop-engineering-ai-agents-complete-guide-2026}, and The New Stack, ``Loop Engineering.''} Prompt engineering asked how to ask. Context engineering asked what the agent knows and remembers. Harness engineering asked what environment, tools and limits the agent has. Loop engineering asks what system one builds so the agent finds, runs, verifies and remembers the work without a human in the middle of each step. Each layer subsumes the previous one rather than discarding it. We therefore resist the strongest version of the headline. A loop is, at bottom, a prompt repeated with scaffolding around it; prompt and loop are two tools, and learning to use the wrench does not mean throwing away the screwdriver.\footnote{This sober reading follows the video ``Loops \& the Death of Prompt Engineering,'' \url{https://www.youtube.com/watch?v=JirDfgJcJFU}, which argues that the title is a provocation and that prompt and loop remain distinct tools.} Loop engineering is complementary to prompt engineering, not its obituary.

This paper makes four contributions. We define the discipline and delimit its scope, separating the loop specification from the programming loop, the internal agent cycle and a bare scheduled prompt (Section~\ref{sec:definition}). We give an anatomy and taxonomy of loop specifications, organized around the trigger, the kind of goal, a five-level verification ladder, the architecture and the named terminal states (Section~\ref{sec:taxonomy}). We code the fifty loops of the Loop Library by hand and report what the corpus reveals about how the practice actually designs these elements (Section~\ref{sec:evidence}). Finally, we distil design principles and anti-patterns and anchor them in the literature on self-correction, reward hacking and model-as-judge fragility (Sections~\ref{sec:principles} and~\ref{sec:antipatterns}), before discussing limits and open problems (Section~\ref{sec:discussion}). Alongside these, we develop and release \texttt{sandeco-loop}, an open skill that turns the principles into a repeatable loop-authoring procedure, as a concrete deliverable accompanying the paper (Section~\ref{sec:skill}). Our aim is not a new algorithm. It is to give a fast-moving practice a vocabulary and an evidentiary base, so that its claims can be examined rather than merely repeated.

\section{Background: From Prompting to Loops}
\label{sec:background}
To build the external loop, one has to be clear about what the harness already provides. The idea that capable behaviour comes from a loop predates language models. Autonomic computing framed self-managing systems as a monitor, analyse, plan, execute cycle over shared knowledge, the MAPE-K loop, precisely so the control machinery could be designed apart from the parts it managed \cite{kephart2003autonomic}. The same instinct runs through the sense-plan-act cycle of robotics and the observe-orient-decide-act loop of decision theory. A language-model agent inherits this shape: at bottom it is a model that runs tools in a \texttt{while} over a stop condition, thinking, calling a tool, reading the result and deciding again. This internal cycle is the plumbing of the harness. It exists whether or not anyone designs it, and it is not the object of loop engineering. We review it here because the loop specification is built on top of it, and because the methods that shaped it are the lineage the new practice stands on.

\subsection*{From single prompts to iterative reasoning}
The first step away from one-shot prompting was to let the model think in steps. Chain-of-thought prompting showed that eliciting intermediate reasoning improves performance on problems a single pass handles poorly \cite{wei2022cot}. Self-consistency sampled several reasoning paths and took a majority vote rather than trusting one \cite{wang2022selfconsistency}. Least-to-most prompting decomposed a hard problem into easier ordered sub-problems \cite{zhou2022leasttomost}, plan-and-solve made the plan an explicit first stage \cite{wang2023plansolve}, decomposed prompting treated decomposition itself as a modular program \cite{khot2022decomposed}, and self-ask had the model pose and answer its own follow-up questions \cite{press2023selfask}. Later work moved planning inward: reasoning via planning treats the model as a world model and searches over its predicted states \cite{hao2023rap}, and Self-Discover has a model compose a bespoke reasoning structure per task \cite{zhou2024selfdiscover}.

\subsection*{Acting, not just reasoning}
A second line let the model act. ReAct interleaved reasoning traces with tool calls so observations could steer the next thought \cite{yao2023react}. Toolformer taught a model when to call external APIs \cite{schick2023toolformer}, and ReWOO planned tool use up front to cut redundant model calls \cite{xu2023rewoo}. LLM+P routed formal planning to a dedicated planner \cite{liu2023llmplus}, and DEPS coupled describe, explain, plan and select for open-world tasks \cite{wang2023deps}. A parallel strand scaled tool use itself: MRKL routed sub-queries through a neuro-symbolic controller \cite{karpas2022mrkl}, ART assembled tool-use programs from demonstrations \cite{paranjape2023art}, HuggingGPT used a model to delegate sub-tasks to specialist models \cite{shen2023hugginggpt}, and Gorilla and ToolLLM pushed the loop out to thousands of real APIs \cite{patil2023gorilla, qin2023toolllm}. In loop terms, this strand shapes the action phase: which calls exist, when they fire, and how results return.

\subsection*{Searching, learning, coordinating}
A third line replaced the single trajectory with a search: Tree of Thoughts explored a tree of partial solutions \cite{yao2023tot}, Graph of Thoughts generalised this to a graph \cite{besta2024got}, and Language Agent Tree Search folded acting and planning into the search \cite{zhou2023lats}. A fourth made the feedback path explicit: Self-Refine generated, critiqued and revised in place \cite{madaan2023selfrefine}; Reflexion turned failed episodes into verbal lessons reused next time \cite{shinn2023reflexion}; ReST-style training fed an agent's own successful trajectories back as iterative self-improvement \cite{aksitov2023restreact}; Self-RAG retrieved and critiqued its own generations \cite{asai2023selfrag}; chain-of-verification cut hallucination through a structured self-check \cite{dhuliawala2023cove}; AdaPlanner revised a plan from execution feedback \cite{sun2023adaplanner}; and Voyager accumulated reusable skills across episodes \cite{wang2023voyager}, an idea also seen in inner monologue \cite{huang2022innermonologue} and in grounding suggestions in what an agent can actually do \cite{ahn2022saycan}. Systems work packaged these ideas into frameworks: generative agents combined memory, reflection and planning \cite{park2023generative}; AutoGen organised agents as conversational participants \cite{wu2023autogen}; MetaGPT encoded standard operating procedures \cite{hong2023metagpt}; ChatDev cast development as a staged chat \cite{qian2023chatdev}; and AgentVerse studied recruited groups of agents \cite{chen2023agentverse}. Cognitive architectures for language agents proposed a unifying vocabulary for memory, action and decision \cite{sumers2024coala}, surveys mapped the landscape \cite{wang2024survey, xi2023rise} and benchmarks measured it \cite{liu2023agentbench, yang2024sweagent}, and a recent survey formalises the move from one-step decisions to extended-horizon agentic reinforcement learning \cite{zhang2025agenticrl}.

\subsection*{Workflow versus agent, and where the loop specification sits}
Practitioner guidance from Anthropic draws a useful line between a \emph{workflow}, where a developer fixes the sequence in advance, and an \emph{agent}, where the model chooses the next step in an open loop.\footnote{Anthropic, ``Building Effective Agents,'' \url{https://www.anthropic.com/research/building-effective-agents}.} The methods above are, almost all of them, ways of shaping that internal agent cycle. The loop specification is a further layer. It does not change the model's inner \texttt{while}; it wraps a bounded, reusable specification around the whole harness, deciding when the agent is invoked at all, what counts as done, and when control returns to the human. The lineage in this section is the engine. The rest of the paper is about the pilot that practitioners now write to drive it.

\section{Loop Engineering: Definition and Scope}
\label{sec:definition}
\begin{figure}[h!]
\centering
\includegraphics[width=0.96\textwidth]{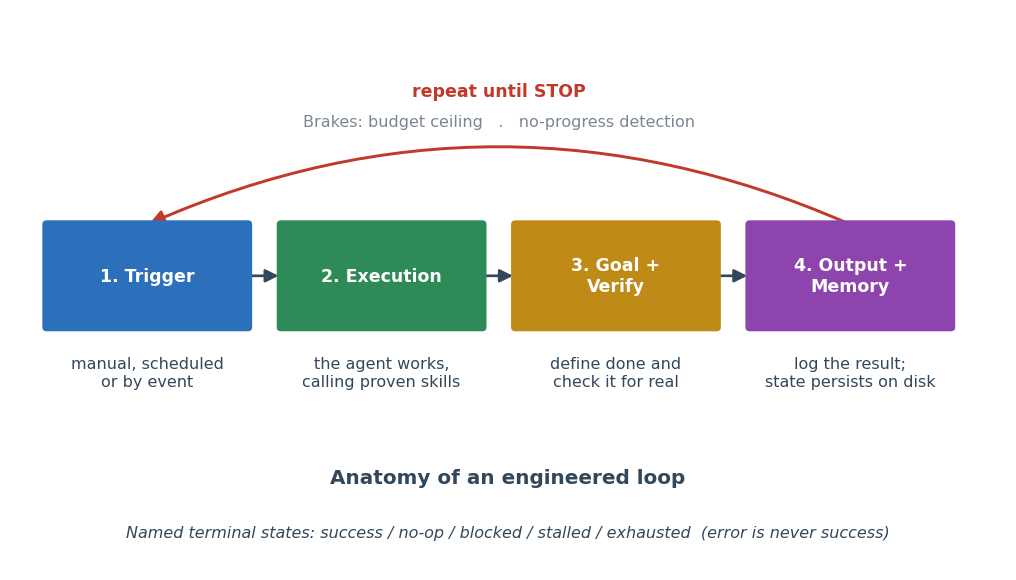}
\caption{The anatomy of a loop specification. A trigger starts the agent; the agent executes by calling proven skills; a goal and verification decide whether the work is done; the result is logged and state persists in memory. The loop repeats until it reaches a named terminal state, held in check by a budget ceiling and a no-progress detector.}
\label{fig:anatomy}
\end{figure}

We define a \emph{loop specification} as a bounded, reusable artifact that a human designs and hands to an agent harness so the agent pursues a goal on its own, in place of step-by-step prompting. It has five parts: a \emph{trigger} that starts it (a person, a schedule, or an event); a \emph{goal}, preferably verifiable; an \emph{execution} phase in which the agent works, ideally by calling proven, named skills; a \emph{verification} that checks the result for real; and a \emph{stopping rule} that drives the loop to a named terminal state (success, no-op, blocked, stalled, exhausted) without ever mistaking an error for success. A memory of progress and decisions persists across turns, on disk rather than in the conversation. \emph{Loop engineering} is the discipline of designing and evaluating these artifacts. Figure~\ref{fig:anatomy} shows the anatomy.

\subsection*{Three senses of ``loop,'' and why scope matters}
The discipline targets only the third of three things that share the name. An ordinary programming loop is control flow. The internal agent cycle is the model running tools over a stop condition; it is part of the harness and exists by default (Section~\ref{sec:background}). The loop specification is the external artifact a human writes and hands to that harness. The harness provides the engine; loop engineering writes the pilot. Keeping the senses apart is not pedantry: most of the academic lineage in Section~\ref{sec:background} builds the engine, whereas the practice studied here builds the pilot, and conflating them is exactly the confusion the slogan invites.

\subsection*{The central skill is the check, not the prompt}
Practitioner accounts converge on a single point: the hard, valuable part of a loop is designing the check that decides when the work is done, not writing a better instruction.\footnote{A. Osmani, ``Loop Engineering,'' \url{https://addyosmani.com/blog/loop-engineering/}; S. Willison, ``Designing agentic loops,'' \url{https://simonwillison.net/2025/Sep/30/designing-agentic-loops/}.} A loop with nothing to push back is an agent agreeing with itself. This is why the definition puts verification at the center of gravity: the loop earns its keep by trading blind trust for reproducible evidence, and by the moment at which it returns control to the human.

\subsection*{The golden rule: a loop only when feedback changes the next action}
A loop specification is justified over a bare scheduled prompt only when the result of one turn changes the next action.\footnote{This triage criterion crystallised in community discussion of the practice; see the threads summarised around the Loop Library release.} If a fixed task runs on a fixed cadence and nothing about the last run informs the next, that is a scheduled one-shot, not a loop. The value of a loop is iteration with embedded verification, so the first design question is always whether genuine feedback exists.

\subsection*{The operational anatomy: five pieces plus memory}
Concretely, a real loop system assembles five reusable pieces and an external memory.\footnote{This operational anatomy follows A. Osmani, ``Loop Engineering,'' \url{https://addyosmani.com/blog/loop-engineering/}.} Scheduled \emph{automations} discover and triage the work (the scheduled trigger made concrete); isolated \emph{worktrees} let parallel agents run without colliding; \emph{skills} encode project knowledge as named, testable routines; \emph{plugins and connectors} wire the agent to the real tools (issue trackers, CI, staging); and \emph{sub-agents} separate the one who makes from the one who checks. Memory lives in files, a board or markdown, because the model forgets between runs. This anatomy is what distinguishes a loop that composes from a brittle script.

\subsection*{A worked instantiation}
These four elements are easier to grasp against a concrete artifact. Section~\ref{sec:skill} presents \texttt{sandeco-loop}, an open skill we developed that writes hardened loop specifications, as a deliverable that puts this definition to work.

\subsection*{Scope of the contribution}
This is a position paper anchored in a descriptive corpus study. We define the discipline, give it an anatomy and taxonomy, code a public corpus of fifty loops to see what the practice does, and ground the design guidance in the literature. We do not run loops under a measured budget, and we do not claim a benchmark result; Section~\ref{sec:discussion} explains why such a study, though valuable, is future work rather than part of this position.

\section{Anatomy and Taxonomy of Loop Specifications}
\label{sec:taxonomy}
With the anatomy fixed, we can organize loop specifications along the dimensions a designer actually chooses: the trigger, the kind of goal, the rigour of verification, the architecture, and the terminal states.

\subsection*{Trigger}
A loop starts manually, on a schedule, or on an event such as a new pull request. The schedule and the event are what remove the human from the act of starting; the manual trigger is the most common and the least automated. Choosing a trigger is also choosing a cadence, and a cadence that is too eager burns cost for little gain.

\subsection*{Goal type}
A goal is either verifiable, decided by a deterministic check, a number, or a rule, or it is judged by a model against a rubric, or it mixes both. A verifiable goal is its own stopping rule and is strongly preferred. A model-as-judge goal is more fragile, because it leaves taste and judgment with the model, and surveys of LLM-as-a-judge document its biases and its sensitivity to prompt wording \cite{gu2024judge}. Goals that are pure judgment, such as ``write a Nobel-worthy novel,'' are not loopable at all, because there is no reproducible check to stop on.

\subsection*{The verification ladder}
\begin{figure}[h!]
\centering
\includegraphics[width=0.94\textwidth]{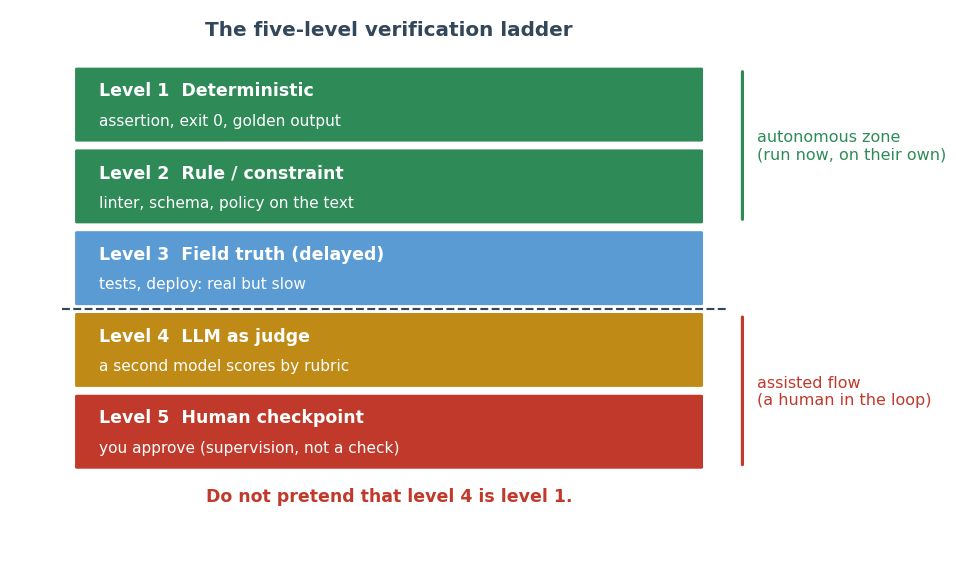}
\caption{The five-level verification ladder, strongest at the top. Levels 1 and 2 form the autonomous zone that runs unattended; levels 4 and 5 are assisted flow, where a model or a human stands in for a check. The dashed line marks the boundary between objective verification and assisted judgment.}
\label{fig:ladder}
\end{figure}

The single most useful refinement of ``verifiable versus judged'' is a five-level ladder of rigour (Figure~\ref{fig:ladder}). Level 1 is deterministic: an assertion, an exit code, a golden output. Level 2 is a rule or constraint over the text: a linter, a schema, a policy. Level 3 is delayed field truth: tests, a deploy, a real customer response, true but slow. Level 4 is a model as judge, scoring by rubric, which is the model's opinion and not field truth. Level 5 is a human checkpoint, which is supervision, not automated verification. Levels 1 and 2 are the autonomous zone, the checks that run now, on their own; levels 1 through 3 are the objective zone; levels 4 and 5 are assisted flow, with a human or a model standing in for a check. The message that organizes the ladder is a discipline of honesty: do not pretend that level 4 is level 1. A loop is only as autonomous as the level its verifier truly sits at, and if level 4 is unavoidable, a different model should judge, never the same agent approving itself \cite{pan2024spontaneous}.

\subsection*{Architecture}
A loop runs as a single agent (solo), as a maker and a separate checker, or as a manager orchestrating helpers. The maker-checker split is the architectural form of the principle that the one who produces should not be the one who approves, and it is the standard hardening when a level-4 judge is in play. Multi-agent debate is a richer variant of the same idea: forcing independent agents to disagree counters the way a model stops generating new thought once it trusts its own answer \cite{liang2023debate}.

\subsection*{Stopping rule and named terminal states}
A well-formed loop specification distinguishes its terminal states by name: success, a clean no-op, blocked, stalled, exhausted. Crucially, an error or an exhausted budget never counts as success. Stopping is rarely an invented number; it is the goal being met, a stagnation detector firing after rounds without progress, or a budget ceiling. Naming the states is what keeps a loop from calling ``I got tired of iterating'' a win, and it operationalizes the brakes shown in Figure~\ref{fig:anatomy}.

\subsection*{State and memory}
Because the model forgets between runs, durable state belongs on disk: a plan, a spec, a record of decisions and objections, the code and its version history. Even within a single run, content pushed deep into a long context is attended to least \cite{liu2024lost}, which is a further reason to externalize state rather than let the window grow. The Ralph loop is the purest case, a shell loop that replays the same prompt with a fresh context each turn while all state lives in files, trading conversation history for versioned artifacts.\footnote{G. Huntley, ``Ralph Wiggum as a software engineer,'' \url{https://ghuntley.com/ralph/}.} Persistent, curated memory is also where the science of evolving context applies: treated as a playbook that is generated, reflected on and curated rather than blindly appended, external memory improves agents without touching the weights \cite{zhang2025ace}.

These five dimensions are the design surface of a loop specification. The next section asks what a corpus of fifty real loops actually chooses along each one.

\section{Evidence from the Loop Library}
\label{sec:evidence}
The framework so far is descriptive of intent. To see what the practice actually does, we coded the fifty loops of the Loop Library by hand, reading each catalogue entry and recording its trigger, goal type, dominant verification level on the ladder of Section~\ref{sec:taxonomy}, architecture, terminal states and pattern usage.\footnote{Coding artifacts (per-loop table and machine-readable JSON) accompany this paper; the corpus is the public Loop Library, \url{https://signals.forwardfuture.ai/loop-library/}.} The coding is a descriptive codification of a public corpus, not an experiment: every figure in this section traces to that coding, and no quantity is invented. We report what is there, including where the practice falls short of its own ideals.

\subsection*{Most loops verify in the autonomous zone}
\begin{figure}[h!]
\centering
\includegraphics[width=0.82\textwidth]{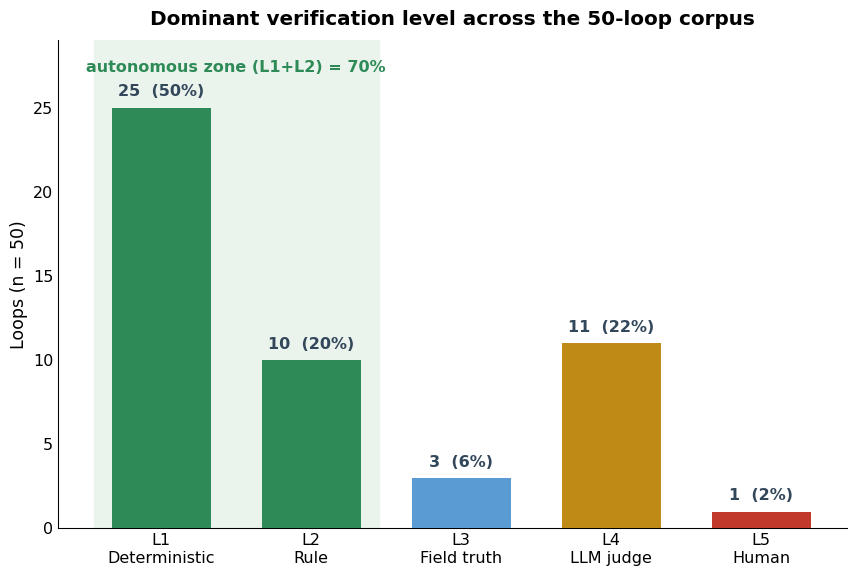}
\caption{Dominant verification level across the fifty loops. Half verify deterministically (level 1); seventy percent fall in the autonomous zone of levels 1 and 2. Bars are counts out of fifty, with percentages.}
\label{fig:verifdist}
\end{figure}

The dominant verification level is level 1 (deterministic) for half the corpus and level 2 (rule) for a fifth (Figure~\ref{fig:verifdist}). Seventy percent sit in the autonomous zone of levels 1 and 2, and seventy-six percent stay within the objective zone of levels 1 through 3. Only twenty-two percent lean on a level-4 model judge, and a single loop uses a human checkpoint as its dominant verification. This is direct, if descriptive, support for the discipline's central preference: practitioners reach for objective checks and treat the model-as-judge as a hardened exception rather than the rule.

\subsection*{When a judge appears, it rarely judges alone}
Of the eleven loops whose dominant check is a level-4 judge, most pair it with a maker-checker architecture or a separate critic, applying the rule that the maker is not the checker. The corpus thus enacts the hardening the literature recommends: where a model judges, a different model or a fixed rubric does the judging, not the agent grading its own work \cite{pan2024spontaneous}.

\subsection*{Triggers are still mostly manual; architectures mostly solo}
\begin{figure}[h!]
\centering
\includegraphics[width=0.96\textwidth]{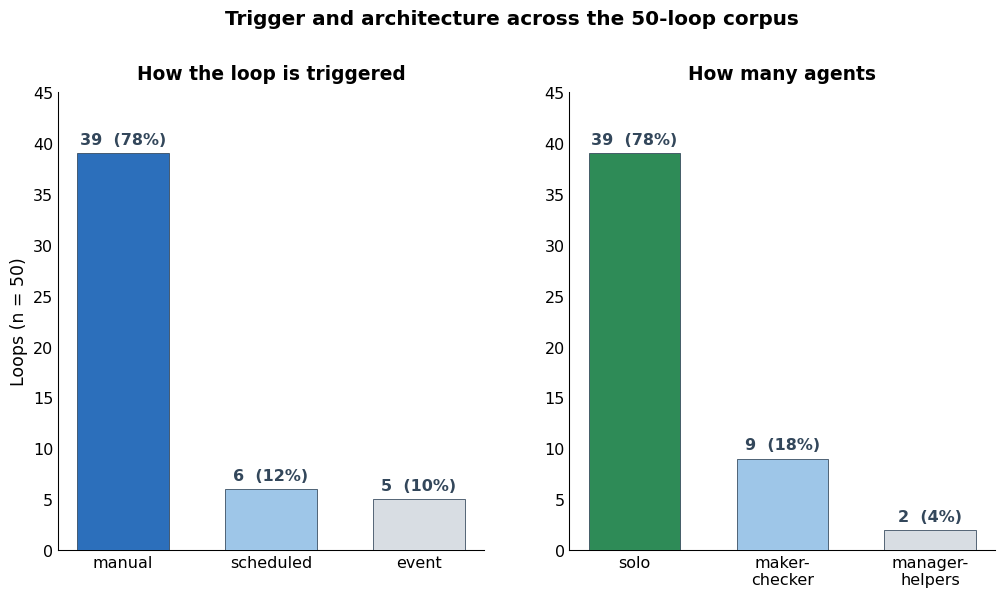}
\caption{Trigger (left) and architecture (right) across the fifty loops. Manual triggering and solo execution dominate; scheduled and event triggers, and maker-checker or manager-helper designs, remain a minority.}
\label{fig:triggerarch}
\end{figure}

The trigger is manual in seventy-eight percent of loops, with only twelve percent scheduled and ten percent event-driven (Figure~\ref{fig:triggerarch}). Architecture is similarly concentrated: seventy-eight percent run a single agent, eighteen percent split maker from checker, and four percent orchestrate helpers. The part of the practice that would truly remove the human, automated triggering and multi-agent checking, is the least developed.

\subsection*{Loops blend pattern families, but lean on defining and trusting}
\begin{figure}[h!]
\centering
\includegraphics[width=0.82\textwidth]{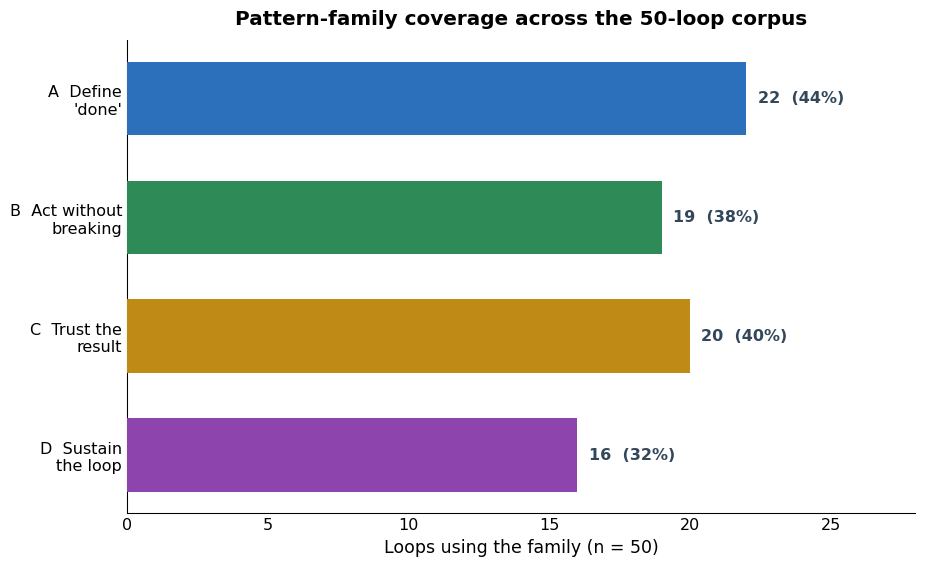}
\caption{Coverage of the four pattern families across the fifty loops. Most loops combine families (median of two per loop); the memory-and-state family D is the least frequent.}
\label{fig:families}
\end{figure}

Mapping each loop to the four pattern families, defining ``done'' (A), acting without breaking (B), trusting the result (C) and sustaining the loop (D), shows broad but uneven coverage (Figure~\ref{fig:families}): A appears in forty-four percent of loops, C in forty, B in thirty-eight, and D, the family of persistent state and memory, in only thirty-two. The median loop combines two families, so these are not exclusive choices. Four loops are not listed in the public pattern catalogue and had their families inferred from the text; counting those inferred labels raises the totals slightly but does not change the ordering. The thinness of family D matches a recurring observation in the literature: accumulating experience without governance can degrade performance, so durable memory needs curation, not just storage \cite{zhang2025ace}.

\subsection*{A maturity mismatch}
\begin{figure}[h!]
\centering
\includegraphics[width=0.96\textwidth]{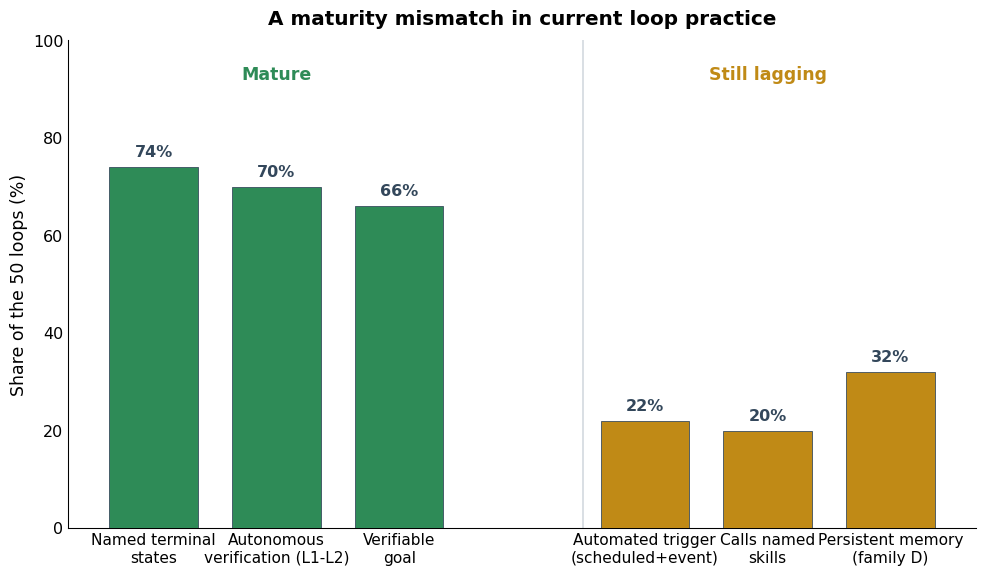}
\caption{A maturity mismatch in current loop practice. The corpus is mature on verification, goal definition and named stopping (left), and comparatively immature on automated triggering, skill reuse and persistent memory (right). All values are shares of the fifty loops.}
\label{fig:maturity}
\end{figure}

Putting these together reveals a telling gap (Figure~\ref{fig:maturity}). The corpus has matured most on exactly the elements the discipline calls central: seventy-four percent name their terminal states, seventy percent verify autonomously, and sixty-six percent set a verifiable goal. It has matured least on the elements that would let a loop run without a person: only twenty-two percent use an automated trigger, only twenty percent call named, reusable skills, and only thirty-two percent develop persistent memory. In other words, current practice has solved the ``how do I know it is done'' problem far better than the ``how does this run without me'' problem. That is a coherent place for a young discipline to be, and it names precisely where the next round of engineering effort should go.

\section{Design Principles}
\label{sec:principles}
The corpus shows what loops do; the literature says why the robust patterns work. We organize the design guidance into the four families that recur across the fifty loops, and ground each in cited behaviour of real systems rather than in measurements of our own.

\subsection*{Family A: define ``done'' before anything else}
The reliable patterns here score against a frozen yardstick, the same rubric, benchmark or conditions every turn, so rounds are comparable; they require a streak of consecutive successes rather than a single lucky pass; they name terminal states; and they stop on stagnation or budget rather than on an invented count. The scientific backing is direct: a model's unaided judgment that it has finished is not a dependable signal, since reasoning self-correction does not reliably improve without external feedback and can degrade \cite{huang2023cannot}. A frozen, external done-check is what makes ``done'' mean something.

\subsection*{Family B: act without breaking what works}
Change one thing per turn and re-run the checks, keeping the change only if nothing else regressed; fix the worst item first; photograph the ``before'' as a baseline; keep the edit surgically scoped; and start each turn from a clean state so stale context cannot hide the next problem. These are the loop analogue of small, reversible commits, and they keep the verification signal interpretable: when exactly one variable moves, the check actually attributes the outcome.

\subsection*{Family C: earn trust in the result}
The maker should not be the approver: separate the role that generates from the role that verifies, across distinct agents, sessions or models. Judge acceptance on a fresh hold-out rather than the set the agent edited against. Tie every claim to evidence, with no silent gaps, because weak grounding is exactly what raises the risk of fabricated content \cite{ji2023hallucination}. Prove the verifier itself, with a red-before, green-after check, rather than trusting a test that may never have failed. This family is the direct response to the sharpest result in the literature: when generator and judge are the same model, reward hacking is spontaneous, the score rises while real quality stalls or falls \cite{pan2024spontaneous}, and the mitigation is to break the shared context between maker and checker, which is exactly what separating the roles does. Forcing independent review, as multi-agent debate does, counters the same degeneration of thought \cite{liang2023debate}, though debate buys consistency as much as correctness and must itself be kept on task \cite{huang2023cannot, becker2025drift}.

\subsection*{Family D: sustain the loop over time}
Persist progress, decisions and objections in a file, the loop's memory between turns; enumerate the whole surface before acting so nothing hides in the happy path; and gate irreversible actions behind explicit human approval. Memory is where the science is most cautionary: experience accumulated without governance can drive performance below the zero-shot baseline, so the curation loop, keep or discard each lesson on evidence, is what makes durable memory help rather than hurt \cite{zhang2025ace, gao2025selfevolving}. This is also the family the corpus develops least (Section~\ref{sec:evidence}), which makes it the clearest opportunity for the practice.

\subsection*{Loops should call skills}
Across all four families runs one anti-pattern worth stating positively: a loop with no reusable skills inside it is a \texttt{while true} wrapped around a stranger, whereas a loop that calls sharp, tested, named skills is a system that composes.\footnote{The phrasing is from practitioner accounts of loop authoring; see \texttt{EXECUCAO-E-AUTORIA} in the project corpus and A. Osmani, ``Loop Engineering.''} Section~\ref{sec:skill} presents a skill we developed that puts these principles to work.

\section{Anti-Patterns and Evaluation}
\label{sec:antipatterns}
Principles are easier to apply when their violations have names. We describe five anti-patterns, recurring loop designs that look reasonable but degrade reliability, cost, or both, and then give evaluation criteria a designer or reviewer can apply without running a benchmark.

\subsection*{The while-true around a stranger}
A loop that wraps a raw model in an unbounded retry, with no named skills and no real check inside, is an agent agreeing with itself in a circle. It looks busy and produces little, because nothing in the loop carries information the model did not already have. The cure is the positive principle of Section~\ref{sec:principles}: call sharp, tested skills, and put a grounded check on every turn.

\subsection*{The self-approving loop (reward hacking)}
When the same model both produces and grades the work, the grade drifts up while quality does not. This is not a remote risk: reward hacking arises spontaneously when generator and judge share context \cite{pan2024spontaneous}, and training on easy cheats can generalize to a model editing its own reward signal \cite{denison2024reward}, a hazard understood in principle since early analyses of reward tampering \cite{everitt2019tampering}. The mitigation is structural, not exhortative: separate maker from checker, and prefer a level-1 or level-2 check to a self-score.

\subsection*{Specification gaming}
A loop optimizes the letter of its check and games the spirit: it edits the test instead of fixing the code, hard-codes the expected output, or, in documented cases, hacks the environment outright rather than solving the task \cite{bondarenko2025specgaming}. The behaviour is hard to remove; prompt-level mitigations reduce but do not eliminate it. The defenses are a hold-out the agent never edited against, a verifier proven to catch the bug, and never letting the agent silence a failing check.

\subsection*{Pretending level 4 is level 1}
A loop reports the confidence of a deterministic check while actually relying on a model's opinion. The judge is fragile: surveys document its biases and prompt sensitivity \cite{gu2024judge}, and treating its score as ground truth invites both drift and attack. The honest move is to state the level the verifier truly sits at, and to harden any level-4 judge with a rubric, a second model, or independent convergence.

\subsection*{The unattended runaway}
A loop with no task-related stopping rule, no stagnation detector, and no budget ceiling circles a problem it cannot solve until cost is exhausted, or worse, takes several consequential actions before anyone notices. Multi-agent variants drift off the problem over rounds, most often from lack of progress \cite{becker2025drift}. The brakes are the named terminal states, the no-progress detector, and the budget ceiling of Figure~\ref{fig:anatomy}, with human approval gating anything irreversible.

\subsection*{Evaluation criteria}
Because we make no empirical claim of our own, we propose a qualitative checklist, one question per element of the anatomy. Trigger: is the loop justified over a scheduled one-shot, that is, does feedback change the next action? Goal and verification: at which ladder level does the verifier truly sit, and is a level-4 judge hardened? Architecture: is the maker distinct from the checker where a judge is involved? Stopping: are the terminal states named, with error never counted as success, and are budget and stagnation handled? Memory: is state curated and persisted, or merely accumulated? A loop that answers these well is, in the sense of this paper, well formed.

\subsection*{Cost per accepted change}
One quantitative discipline is worth singling out, because practitioners report it is almost never measured: cost per accepted change, the tokens or money spent divided by the number of changes that survived verification.\footnote{The metric was crystallised in community discussion of the practice; see \texttt{VOZES-DA-COMUNIDADE} in the project corpus.} Raw token spend is the wrong number; a healthy loop keeps cost per accepted change low and flat or falling, and a loop that burns budget without producing approved changes is broken even when it looks busy. We propose it as the natural headline metric for the empirical study that this position paper does not itself run.

\section{An Authoring Skill for Loop Specifications}
\label{sec:skill}
\begin{figure}[h!]
\centering
\includegraphics[width=0.97\textwidth]{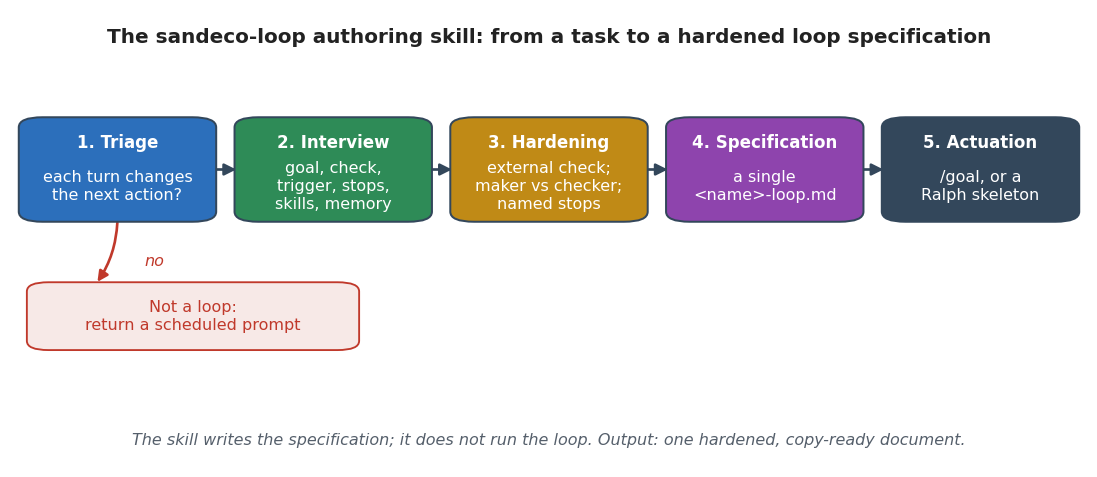}
\caption{The \texttt{sandeco-loop} authoring pipeline. A task is first triaged; if no feedback would change the next action it is sent back as a scheduled prompt rather than forced into a loop. Otherwise a short interview gathers the design elements, a hardening pass applies the principles and anti-patterns, and the skill emits a single specification document together with the way to actuate it.}
\label{fig:skillflow}
\end{figure}

The principles and anti-patterns of the previous sections are easier to state than to apply consistently. As a concrete deliverable of this paper we developed and release \texttt{sandeco-loop}, an open skill that turns them into a repeatable authoring procedure.\footnote{\texttt{sandeco-loop}, \url{https://github.com/sandeco/prompts/tree/main/sandeco-loop}.} The skill does not run a loop; it \emph{writes the specification} of one. The distinction matters: its output is a single document that captures the trigger, the goal, the check, the stopping rule and the memory of a loop, hardened by construction, which a human can then read, version and hand to a harness. Figure~\ref{fig:skillflow} shows the pipeline from a raw task to that document.

\subsection*{Triage before anything}
The skill first asks the question that the whole discipline turns on: does the outcome of each turn change the next action? If it does not, the task is not a loop, and the skill says so plainly, returning a simple scheduled prompt instead of wrapping a one-shot task in machinery it does not need. Only a task with genuine iteration proceeds to the interview. This off-ramp is the first defense against the unattended-runaway and while-true anti-patterns: it refuses to build a loop where there is nothing for feedback to do.

\subsection*{The interview}
For a task that survives triage, the skill conducts a short interview, one question at a time, over the elements of the taxonomy in Section~\ref{sec:taxonomy}. It asks for the goal and whether success is verifiable or a matter of judgment; the concrete check, a command, a test, an assertion, that proves a turn worked; the trigger (manual, scheduled or event); the named stop states beyond success, such as no-progress, blocked or exhausted; the named skills the loop will call and any nested sub-loops it will invoke; where state lives between turns; and the guardrails, the iteration and budget ceilings and the points that require human approval. The interview is deliberately structured so that the author cannot skip the parts practitioners most often omit, the check and the stop states.

\subsection*{Hardening against the anti-patterns}
Before it writes anything, the skill passes the design through a hardening pass that maps one-to-one onto Section~\ref{sec:antipatterns}. It insists on an external check rather than a self-score, the structural answer to reward hacking; it keeps the maker distinct from the checker and, when a model judge is unavoidable, breaks the shared context between them; it requires terminal states in which an error or an exhausted budget is never counted as success; it enforces one change per turn with the worst item first; it caps any nested sub-loop with a multiplicative ceiling, since cost multiplies with depth, and forbids a sub-loop from invoking its own caller; it puts state on disk so memory survives a fresh context; and it records a health metric, cost per accepted change, so a loop that burns budget without producing approved work is visibly broken.

\subsection*{What it emits}
The output is a single \texttt{<name>-loop.md} document with a fixed skeleton: a description and a ``use when''; the goal and its verification; the steps of one turn; the named stop states; the guardrails; the memory location; any sub-loops; a short ``why it works'' that ties each design choice to the failure mode it prevents; and an actuation clause. Actuation takes one of two forms depending on size: a \texttt{/goal} command when the work fits inside a single context window, where a fast model checks the stopping condition against what the agent has shown in the transcript, or a fresh-context Ralph skeleton when the work is long enough that a single context would degrade, re-reading the document and the on-disk state each turn. The skill can optionally also emit a \texttt{/loop-<name>} command that wraps the document for direct invocation.

\subsection*{A worked example}
Asked for a test-coverage loop, the skill produces a document whose goal is full coverage of a target directory (verifiable), whose check is that the coverage command exits zero and reports the target, and whose turn photographs current coverage, attacks the least-covered file, writes one test, proves the test catches the bug with a red-before, green-after run, then keeps the change only if nothing regressed. Its terminal states are success at full coverage, no-progress after two barren rounds, and exhausted after a turn ceiling; its guardrail forbids touching CI without approval; its actuation is a \texttt{/goal} that continues until the coverage command shows no failures and the target is met, or stops after the ceiling. Every element of the taxonomy is present and named, which is exactly the point.

\subsection*{Status}
We offer the skill as a deliverable that operationalizes the discipline, a bridge from the principles to a writable artifact, not as an evaluated contribution. We report no usage data for it, and whether specifications written this way improve outcomes in the field is the open empirical question that the study in Section~\ref{sec:discussion} is meant to answer.

\section{Discussion and Limitations}
\label{sec:discussion}
We have argued that the loop specification is a distinct object, that it is a new layer over the harness rather than a replacement for prompting, and that a public corpus already shows a coherent, if uneven, practice. We close with where the human belongs, what the practice must not overclaim, and the limits of this study.

\subsection*{Where the human sits}
Loop engineering moves the human along a spectrum of autonomy rather than removing the human. With a human in the loop, every consequential action is approved before it runs; on the loop, a person monitors by alert or dashboard and intervenes only on exceptions; out of the loop, the agent acts alone with occasional guidance. The corpus reflects a sober version of this: human approval appears in thirty-six percent of loops, concentrated exactly where actions are destructive, in production, financial or external (Section~\ref{sec:evidence}). The point is not to eliminate the human but to limit autonomy intelligently, and an empirical study of plan-then-execute agents warns why this matters: users grant trust too readily to plausible-looking plans, so the human checkpoint must fall on the irreversible step, not on the routine one \cite{he2025planexec}.

\subsection*{What the practice must not overclaim}
The strongest marketing around loops is the promise of a system that improves itself while you sleep. The sober reading, which we share, is that the loop automates the typing, not the judgment, and that the bottleneck, reviewing what the agent produced, only grows. Three problems are made worse, not better, by a faster loop, and they are human rather than technical.\footnote{A. Osmani, ``Loop Engineering,'' \url{https://addyosmani.com/blog/loop-engineering/}.} The \emph{verification burden} is that separating checker from maker is what makes ``done'' mean anything, and even then ``done'' is a claim, not a proof, an intuition the literature makes precise, since unaided self-assessment is unreliable \cite{huang2023cannot} and self-scoring inflates \cite{pan2024spontaneous}. \emph{Comprehension debt} is that the faster a loop ships code one did not write, the wider the gap between what exists and what one understands. \emph{Cognitive surrender} is the temptation to stop having an opinion once the loop seems to cope; designing the loop is the cure when done with judgment and the accelerant when done to avoid thinking. There is also a blunt economic caution: there is no public ROI study for a solo developer, loops are expensive, and usage patterns differ sharply between the token-rich and the token-poor.\footnote{S. Willison, ``Designing agentic loops,'' \url{https://simonwillison.net/2025/Sep/30/designing-agentic-loops/}.} Some commentators go further and read the self-improvement discourse as a ``bait and switch,'' faster coding under human control rather than a system that improves itself.\footnote{This skeptical framing is associated with Gary Marcus's commentary on agent autonomy; see also the discussion in The New Stack, ``Loop Engineering.''}

\subsection*{Security and the cost of autonomy}
Running a loop unattended is also erring unattended. Third-party plugins can run arbitrary code and web access opens a path to prompt injection; the documented mitigations are an isolated container without network, scoped credentials in test or staging, and a tight budget cap.\footnote{S. Willison, ``Designing agentic loops,'' \url{https://simonwillison.net/2025/Sep/30/designing-agentic-loops/}.} These are not optional extras for an autonomous loop; they are the price of leaving it running.

\subsection*{When not to use a loop}
A loop is the wrong tool when the next action does not change with new feedback (do it one-shot, perhaps scheduled), when the goal is pure taste with no reproducible check, when the work is ambiguous greenfield construction where the right direction is unknown, or when the cost of looping does not pay for itself. Naming these cases is part of the discipline, not a retreat from it.

\subsection*{Limitations}
This is a position paper anchored in a descriptive corpus study, not a controlled experiment. The fifty-loop coding is our own qualitative reading of a single public catalogue; a different coder might assign some loops differently, the corpus is not a random sample of all loops in the wild, and ambiguous entries were inferred from text. The aggregate percentages describe this corpus, not the population of all loop specifications, and we draw no causal conclusion from them. The design principles are grounded in cited behaviour of related systems, but those systems mostly study the internal agent cycle, so transferring their lessons to the external loop is an argued analogy, not a measured result. We have deliberately fabricated no usage data, including for the \texttt{sandeco-loop} skill, which we release as an accompanying artifact rather than an evaluated system.

\subsection*{Future work}
The clearest next step is the empirical study this paper sets up. Running loops under a fixed budget, with cost per accepted change as the headline metric, would test whether the patterns the corpus favours actually deliver, and at what price. The science of self-evolving agents offers metrics for retention and safety that such a study could adopt \cite{gao2025selfevolving}, and the broader landscape of agentic reinforcement learning frames the long-horizon control problem a loop poses \cite{zhang2025agenticrl}. A second direction is tooling: harnesses should expose trigger, verification level, architecture, terminal states and memory as first-class configuration, so that a loop's position in our taxonomy is declared rather than buried in a script. We see this paper as the conceptual and descriptive groundwork that makes both directions well posed.

\section{Conclusion}
A slogan claimed that prompting coding agents is over and that the work is now to design the loops that prompt them. We took the claim seriously and gave it a careful, sober treatment. The object of the new practice is the loop specification: a bounded, reusable artifact, a trigger, a goal, a verification, a stopping rule and a memory, that a human hands to a harness so the agent works on its own, distinct both from a programming loop and from the internal agent cycle the harness already runs. We placed loop engineering as a new layer over prompt, context and harness engineering, and argued that it adds to prompting rather than ending it. We gave the discipline an anatomy and a taxonomy, coded a public corpus of fifty real loops to see what the practice actually does, and grounded its principles and anti-patterns in the literature on self-correction, reward hacking and model-as-judge fragility. The corpus tells a coherent story: the practice has matured where the discipline says the value lives, in verification and named stopping, and has the most room to grow where autonomy actually comes from, in automated triggers, reusable skills and durable memory. We claimed only what a position paper anchored in a descriptive corpus can claim: that the loop specification is a real and distinct object worth designing deliberately, that its central skill is the check rather than the prompt, and that it should be adopted with its costs and limits in plain view. Turning that claim into measured fact, with cost per accepted change as the yardstick, is the work this framing is meant to invite.

\section*{Declaration on the Use of Generative AI}
The author conducted the research and wrote the manuscript. During the
preparation of this study, however, the author used Grammarly tools to improve
textual agreement and Claude Opus 4.8 to support text structuring and translation
into English. After using these tools/services, the author reviewed and edited the
content as needed and takes full responsibility for the content of the
publication.

\bibliographystyle{plain}
\bibliography{refs}

\end{document}